\documentclass[%
reprint,
amsmath,amssymb,
aps
]{revtex4-1}

\usepackage{graphicx}
\usepackage{dcolumn}
\usepackage{bm}

\begin{document}

\title{Topological symmetry classes for non-Hermitian models and connections to the bosonic Bogoliubov-de Gennes equation}

\author{Simon Lieu} 

\affiliation{
   Blackett Laboratory, Imperial College London, London SW7 2AZ, United Kingdom
   }

\date{\today}
\begin{abstract}
The Bernard-LeClair (BL) symmetry classes generalize the ten-fold way classes in the absence of Hermiticity. Within the BL scheme, time-reversal and particle-hole come in two flavors, and ``pseudo-Hermiticity"  generalizes Hermiticity. We propose that these symmetries  are relevant for the topological classification of non-Hermitian single-particle Hamiltonians and \textit{Hermitian} bosonic Bogoliubov-de Gennes (BdG) models. We show that the spectrum of \textit{any} Hermitian bosonic BdG Hamiltonian is found by solving for the eigenvalues of a non-Hermitian matrix which belongs to one of the BL classes. We therefore suggest that bosonic BdG Hamiltonians  inherit the topological properties of a non-Hermitian symmetry class and explore the consequences by studying symmetry-protected edge instabilities in a simple 1D system.
\end{abstract}

\maketitle

\section{Introduction}

Topological phase transitions were first discovered in gapped fermionic systems, described by quantum Hamiltonians which are necessarily Hermitian operators \cite{Qi1,Hasan1}. More recent developments have revealed that topological transitions are  ubiquitous and extend well beyond the world of closed fermionic models. Photonic crystals \cite{Weimann1, Poli1,Peng1}, classical plasmonic chains \cite{Pocock1,Downing1},  and lasers \cite{Amo1, Malzard1} exhibit hallmark characteristics of topology: robust edge modes, quantized physical observables, bulk invariants. Often these systems can be described via a non-Hermitian Hamiltonian, with the real and imaginary parts of the eigenvalues related to the frequency and decay rate of the modes. Additionally, recent proposals suggest that such Hamiltonians can be used to model the finite lifetime of quasiparticles in closed fermionic systems with disorder and/or interactions \cite{Kozii1}. A growing body of literature has thus been dedicated to the study of topological phase transitions in the absence of Hermiticity \cite{Shen1, Gong1,Lee1, Zhou1,Kunst1,Yao1,Yin1,Wunner1,Lieu}.

Quadratic fermion Hamiltonians can be sorted into ten classes \cite{Altland1} which are based on three symmetries: time-reversal, particle-hole, and chiral. These ten classes form the basis of the periodic table of topological insulators/superconductors which guarantees a topological or trivial classification for each class in a given dimension, known as the ten-fold way \cite{Ryu1}. A similar symmetry-based framework which encapsulates all non-Hermitian models is currently lacking. In this work we begin to make progress in this direction by highlighting a set of symmetry classes which were originally introduced by Bernard and LeClair (BL) as non-Hermitian generalizations of the ten Altland-Zirnbauer random matrix classes \cite{Bernard1}. An important work by Esaki \textit{et al} \cite{Esaki1} found that these symmetries are very relevant when discussing topological transitions,  since they include uniquely non-Hermitian relationships which protect edge modes.

Surprisingly, the BL classes are also relevant for \textit{Hermitian} bosonic models. While topological fermionic BdG models (i.e. superconductors) are fully understood within the ten-fold way, no such classification exists for bosonic BdG models yet. Hermitian bosonic BdG models are diagonalized via a symplectic (non-unitary) transformation. We show that the resulting spectrum is equivalent to the eigenvalues of a non-Hermitian matrix and, moreover, that this matrix belongs to one of the BL classes. Remarkably, it is thus  useful to consider non-Hermitian random matrix classes when discussing Hermitian bosonic BdG models.  While typical equilibrium setups have a spectrum which is fully real \cite{Chalker1}, modes with ``imaginary energies" can arise as dynamical instabilities of a quenched bosonic system \cite{Ketterle1, Demler1, Nakamura1,Barnett1,Barnett2}.  We demonstrate this explicitly by studying a 1D bosonic Su-Schrieffer-Heeger (SSH) model \cite{Su1,Barnett1} and examining the robustness of symmetry-protected edge instabilities.  The BL framework will thus allow us to discuss non-Hermitian single-particle models and topological bosonic BdGs in the same language. Our analysis thus provides a link between the study of general non-Hermitian topological transitions and topological bosons.

\section{Motivation to expand the ten-fold way}

We briefly review the ten Hermitian symmetry classes \cite{Altland1} before motivating the need to consider alternative symmetries when discussing topological transitions in non-Hermitian models. Consider a matrix $H$ which represents a first quantized Hamiltonian, e.g.  $\mathcal{H}=\sum_{i,j} H_{i,j} c_i^\dagger c_j$. For any $H$ we can check for the presence of the following  symmetries:
\begin{align}
H & =x H^{*}x^{\dagger},\qquad x x^{*}=\pm\mathbb{I}\label{eq:x}\\
H & =-y H^{*}y^{\dagger},\qquad y y^{*}=\pm\mathbb{I}\label{eq:y}\\
H & =-z Hz^{\dagger},\qquad z^{2} = \mathbb{I}\label{eq:z}
\end{align}
which are known as time-reversal, particle-hole, and chiral symmetry respectively. One can construct ten unique classes based on the presence or absence of these symmetries. These ten classes are of paramount importance since it has been shown that each symmetry class can either possess $\mathbb{Z},\mathbb{Z}_2,$ or 0  topologically distinct phases in a given dimension $d$ for Hermitian, non-interacting models \cite{Ryu1}.  How much of this analysis survives the breaking of Hermiticity?

Consider the dissipative bipartite quantum-walk model in 1D:
\begin{align}
\mathcal{H}_{\text{qw}} =  v \sum_{i=1}^{m}  \left(  c_{A,i}^{\dagger} c_{B,i}  + h.c.\right)  &+  w \sum_{i=1}^{m-1}  \left(c_{B,i}^{\dagger} c_{A,i+1}  + h.c.\right) \nonumber \\  &-iu \sum_{i=1}^{m}   c_{B,i}^{\dagger}  c_{B,i}
\end{align}
where $u,v,w \in \mathbb{R}$ and $c_{A/B,i}$ represents an annihilation operator on lattice site $i$ in sublattice site $A/B$.  \cite{Rudner1}. Physically, this represents a standard SSH model with dissipation on one of the sublattice sites only, and has been realized in an optical setting \cite{Rudner2}. The first-quantized Hamiltonian matrix $H_{\text{qw}}$  does \textit{not} have time-reversal, particle-hole, or chiral symmetry and therefore belongs to class A. The Hermitian ten-fold way analysis would suggest that this model ought to be trivial in 1D.  However,  Esaki \textit{et al} \cite{Esaki1} found that this model hosts strongly protected edge modes \footnote{The model studied was slightly different from  the dissipative quantum walk, however the argument remains valid.}. While chiral symmetry is broken, edge modes are protected by an inherently non-Hermitian relationship:  $H_{\text{qw}}=-\tau_{z} H_{\text{qw}}^{\dagger}\tau_{z}$, $\tau_{z}= \mathbb{I}_N \otimes \sigma_z$, ($\sigma_i$ represents Pauli matrices) which constrains edge mode energies via $E_{\text{edge}} = -E_{\text{edge}}^* $. Any (gap preserving) disorder entering the Hamiltonian which respects this relationship will ensure that edge modes will have exactly zero real energy, $\text{Re}[ E_{\text{edge}}]=0$ (though their imaginary energy can be arbitrary). This implies that non-Hermitian models can possess edge-mode-protecting symmetries which are not included in the ten-fold way. Classifying symmetries should be sensitive to more general relationships between $H$ and $H^\dagger$.

A recent study has analytically derived the topological periodic table of the ten-fold way in the absence of Hermiticity \cite{Gong1}. An interesting prediction of this work is that  all ten classes are topological in 1D,  while all are trivial in 2D. This analysis again suggests that symmetries beyond the ten-fold way ought to be taken into account to potentially achieve distinguishability between topological and trivial phases in 1D and 2D. 

Guided by this intuition, in the next section we identify a more comprehensive  classification (at the cost of more symmetry classes) for  non-Hermitian models. 

\section{Bernard-L\MakeLowercase{e}Clair symmetry classes }

We describe the symmetries which define the BL classes \cite{Bernard1}. Any first-quantized Hamiltonian matrix $H$ will be sorted based on the presence or absence of the following relations:
\begin{align}
H & =-p Hp^{\dagger},\qquad p^{2}=\mathbb{I}\label{eq:p}\\
H & =\epsilon_{c}c H^{T}c^{\dagger},\qquad c c^{*}=\pm\mathbb{I}\label{eq:c}\\
H & =\epsilon_{k}k H^{*}k^{\dagger},\qquad k k^{*}=\pm\mathbb{I}\label{eq:k}\\
H & =\epsilon_{q}q H^{\dagger}q^{\dagger},\qquad q^2=\mathbb{I}\label{eq:q}
\end{align}
where $\epsilon_{c,q,k}=\pm1$ and $\mathbb{I}$ is the identity. We mandate that the symmetries commute and are of order two (i.e. each symmetry operation squares to the identity) which implies the constraints on the right.  We label these symmetries as $P,C,K,Q$ respectively.

The physical motivation to consider these relationships is as follows. The $P$ symmetry is standard chiral symmetry. The $C,K$ symmetries can be viewed as two distinct flavors of particle-hole ($\epsilon=-1$) or time-reversal ($\epsilon=+1$). This is because in the Hermitian case $H^T=H^*$ however $H^T$ differs from $H^*$  in the non-Hermitian case. Therefore the $C,P$ symmetries along with Hermiticity exhaust the standard ten-fold way. In addition, we will consider $Q$ symmetry (or ``pseudo-Hermiticity") which generalizes Hermiticity.

BL found that these four relations result in 43 distinct symmetry classes which may be viewed as non-Hermitian generalizations of the ten-fold way classes. For details we refer to Refs.~\cite{Bernard1,Magnea1}.

Having introduced these relationships and discussed some of their basic properties, we will now show that any bosonic BdG equation will possess a $Q$  and a $K$ symmetry by construction. The BL classification thus provides us with a unified framework to discuss both non-Hermitian single-particle Hamiltonians and Hermitian bosonic BdG models simultaneously. 

\section{pseudo-Hermiticity in the general bosonic B\MakeLowercase{d}G equation}

We demonstrate that the spectrum of every Hermitian bosonic BdG Hamiltonian is equivalent to the eigenvalues of a non-Hermitian matrix. We then sort this matrix into one of the BL classes.

Let us consider a  Hermitian $2N\times2N$ bosonic BdG Hamiltonian
\begin{equation}
\mathcal{H}=\mathbf{b}^{\dagger}\tilde{H}\mathbf{b},\qquad\mathbf{b}=\left(b_{1},\ldots,b_{N},b_{1}^{\dagger},\ldots,b_{N}^{\dagger}\right)^{T}
\label{eq:general}
\end{equation}
where $b$ are bosonic operators and
\begin{equation}
\tilde{H}=\left(\begin{array}{cc}
B & A\\
A^{*} & B^{T}
\end{array}\right)
\end{equation}
represents the most general form where $B=B^\dagger, A=A^T, \tilde{H}=\tilde{H}^{\dagger}$. We define the transformation
\begin{equation}
\mathbf{b}=T\mathbf{\beta},\qquad\mathbf{b}^{\dagger}=\mathbf{\beta}^{\dagger}T^{\dagger},\qquad T=\left(\begin{array}{cc}
U & V\\
V^{*} & U^{*}
\end{array}\right)
\end{equation}
such that 
\begin{equation}
\mathcal{H}=\mathbf{\beta}^{\dagger}\Sigma_{z}\Lambda\mathbf{\beta}
\end{equation}
where $\Sigma_{z}=\sigma_{z}\otimes\mathbb{I}_{N}$ and $\Lambda$ is a diagonal matrix. If we impose bosonic commutation relations on the quasiparticles $\beta$ then the transformation $T$ must obey
\begin{equation}
T\Sigma_{z}T^{\dagger}=\Sigma_{z}\Rightarrow T^{-1}=\Sigma_{z}T^{\dagger}\Sigma_{z} \label{eq:transform}
\end{equation}
which implies $T$ is \textit{symplectic} (i.e. non-unitary). In addition, $T$ needs to diagonalize the Hamiltonian
\begin{equation}
T^{\dagger}\tilde{H}T=\Sigma_{z}\Lambda\Rightarrow\tilde{H}T=\left(T^{\dagger}\right)^{-1}\Sigma_{z}\Lambda=\Sigma_{z}T\Lambda.
\end{equation}
Multiplying by $\Sigma_{z}$ on the left leads to
\begin{equation}
\Sigma_{z}\tilde{H}T=T\Lambda
\label{eq:bdgeqn}
\end{equation}
which implies that the eigenvalues of $H_{\text{BdG}}\equiv\Sigma_{z}\tilde{H}$ will determine the  spectrum of the system, and the right-eigenvectors will determine the transformation matrix $T$. We refer to \eqref{eq:bdgeqn} as the bosonic BdG equation, and point out two symmetries built into $H_{\text{BdG}}$:
\begin{align}
H_{\text{BdG}}&=\Sigma_{z}H_{\text{BdG}}^{\dagger}\Sigma_{z}\label{eq:bdgq} \\
H_{\text{BdG}}&=-\Sigma_{x}H_{\text{BdG}}^*\Sigma_{x}\label{eq:bdgk}
\end{align}
corresponding to a $Q$ symmetry with $\epsilon_{q}=+1$ (pseudo-Hermiticity) and a $K$ symmetry with $\epsilon_{k}=-1,kk^*=\mathbb{I}$ (particle-hole) \cite{Nakamura1}. This class is also defined by: $q=-k q^* k^\dagger$ from the commutation condition, where the minus sign distinguishes it from the Hermitian class D. Thus while the Hamiltonian \eqref{eq:general} is Hermitian, the spectrum is equivalent to the eigenvalues of a non-Hermitian matrix. 

Note that this analysis does not apply to fermionic BdG models (i.e. superconductors). These are diagonalized by a unitary transformation such that the fermionic BdG equation is particle-hole symmetric and Hermitian: $\epsilon_{q}=+1, q=\mathbb{I}$, $q=+k q^* k^\dagger$, representing class D \cite{Altland1}. 

The non-Hermiticity of $H_{\text{BdG}}$ suggests that imaginary modes can appear which are populated exponentially fast in time (also known as dynamically-unstable modes). Physically, these modes can arise if bosons condense in a state which does not minimize the mean-field energy of the evolving Hamiltonian, such as after a quench protocol. The system proceeds to rapidly depopulate the condensate into one of the unstable modes, which has been observed experimentally in cold atoms \cite{Ketterle1}. This condensate depletion mechanism is absent from fermionic BdG models. This is the physical reason why bosonic and fermionic BdGs belong to separate symmetry classes.

\section{A bosonic-SSH model}

Up to now our discussion has been general. To explicitly show the effects of non-Hermiticity in the bosonic BdG equation, we will study the simplest 1D model which undergoes a topological transition and observe symmetry-protected edge instabilities.

Consider the Hamiltonian
\begin{multline}
\mathcal{H}_{b}=  v \sum_{i=1}^{n}  \left(  b_{A,i}^{\dagger} b_{B,i}  + h.c.\right) +  w \sum_{i=1}^{n-1}  \left(b_{B,i}^{\dagger} b_{A,i+1}  + h.c.\right)\\ +  \sum_{i=1}^{n} \left[ u \left(  b_{A,i} b_{A,i} +  b_{B,i}  b_{B,i}\right) +h.c. \right] \label{eq:bdgssh}
\end{multline}
where $v,w\in \mathbb{R}$, $u\in \mathbb{C}$, and $b_{A/B,i}$ represents a bosonic operator on lattice site $i$ in sublattice site $A/B$.  This Hamiltonian was proposed in Ref.~\cite{Barnett1} to describe 1D bosons initially prepared in a higher-energy band then allowed to evolve.  Note that the Hamiltonian $\mathcal{H}_{b}$ is Hermitian, however (due to the arguments in the previous section) the matrix which we need to diagonalize in order to obtain the spectrum ($H_{\text{BdG}}$) is non-Hermitian. Within the BL framework, if $u\in \mathbb{C}$ then the only two relevant symmetries of the model are given in Eqs.~(\ref{eq:bdgq}) and (\ref{eq:bdgk}), namely a $Q$ and a $K$ symmetry.

\begin{figure}[b]
\begin{centering}
\includegraphics[scale=0.1]{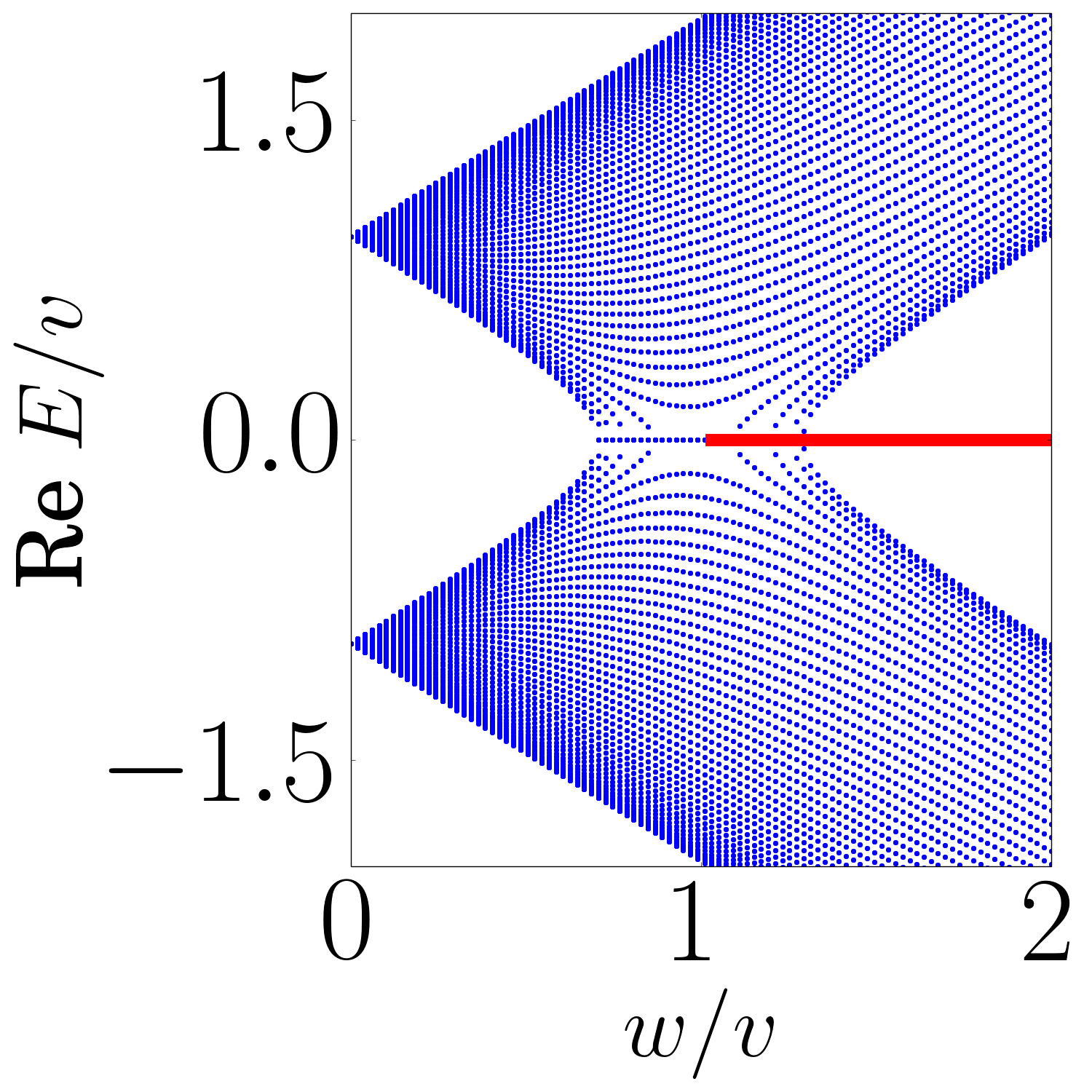}
\includegraphics[scale=0.1]{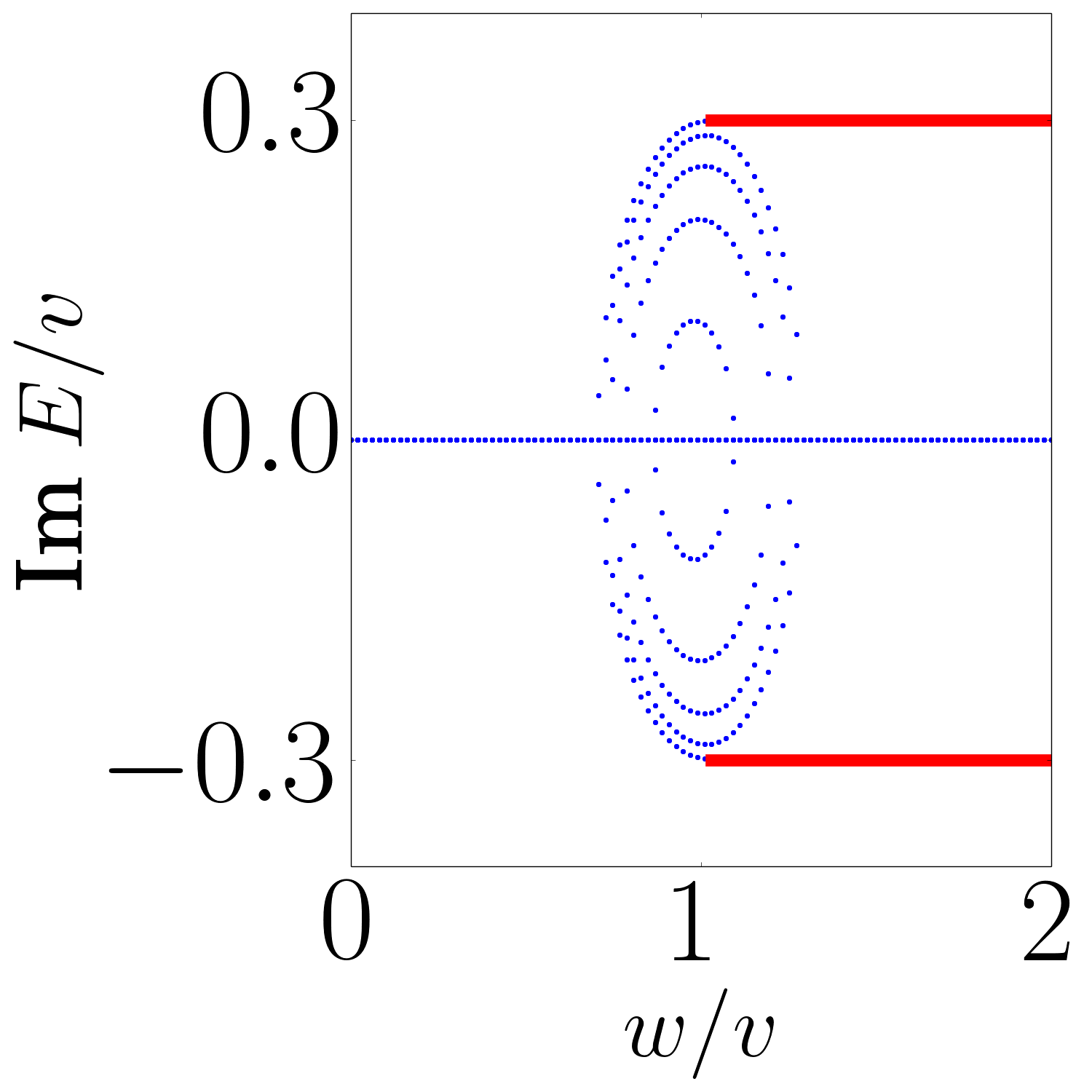}
\par\end{centering}
\caption{\label{fig:bdgSpec} Bosonic-SSH post-quench spectrum with $u/v =0.3\exp(i\pi/5),n=100$. Edge modes given in red.}
\end{figure}

The spectrum with open boundary conditions is given in Fig.~\ref{fig:bdgSpec} as a function of $w/v$ at a constant value of $u/v$. The bulk spectrum is purely real and gapped for parameters $|w-v|>|u|$, however away from this regime some bulk modes start to become unstable and come in complex conjugate pairs. This is known as a broken-to-unbroken transition, discussed in the Appendix. 

Of particular interest are the edge modes (given in red), which have a purely imaginary energy. Each edge hosts two modes: one with energy $+i |u|$ and one with $-i |u|$. Edge modes are particle-hole symmetric: $ t_{\text{edge},+}  \propto \Sigma_x  t_{\text{edge},+}^*$ which suggests that $E_{\text{edge}, +} = -E_{\text{edge}, +}^*$ where $t_{\text{edge},+} $ represents an eigenstate of $H_{\text{BdG}}$ on an edge  with energy $E_{\text{edge}, +}=+i |u|$. Thus because the mode on the edge is its own particle-hole symmetric partner, its energy must be purely imaginary. The stability of these edge modes will be discussed in the next section.

Since we have found that edge modes are particle-hole symmetric, one may wonder whether edge excitations are inherently \textit{non-local}, in analogy with Majorana excitations in a Kitaev chain \cite{Kitaev1}. This interpretation is blurred when discussing the aforementioned model, since the Hamiltonian (\ref{eq:bdgssh}) cannot be brought to fully diagonal form in terms of bosonic quasiparticles when associated energies are imaginary. This is because complex modes are generically non-normalizable, i.e. $t_{\text{edge},+}^\dagger \Sigma_z t_{\text{edge},+}=0$  \cite{Nakamura1,Takahashi1} which implies that the Bogoliubov transformation (\ref{eq:transform}) cannot be satisfied. There is no constraint for edge excitations to be represented as non-self-adjoint bosonic quasiparticles. While the Hamiltonian is not diagonalizable via bosons, the imaginary eigenvalues of $H_{\text{BdG}}$ still have physical meaning: the Heisenberg equations of motion reveal exponential increase in edge site population as a function of time \cite{Barnett1}.

\section{Robustness of bosonic-SSH edge modes} \label{stability}

We return to the question of the robustness of edge modes in the bosonic-SSH model via an adiabatic argument. A few subtleties arise due to the non-Hermiticity of $H_{\text{BdG}}$.  We begin by considering a parameter regime where all the bulk modes are fully real and gapped, and we restrict our attention to the left edge for concreteness.  First, recall that the this edge hosts two right-eigenvectors of $H_{\text{BdG}}$: $t_{\pm}$ with associated energies $E_{\pm}=\pm i |u|$ respectively. These modes are particle-hole symmetric, satisfying $ t_{\pm}  \propto \Sigma_x  t_{\pm}^*$ which constrains energies via $E_{\pm}=-E_{\pm}^*$. For Hermitian models, this condition results in the modes having exactly zero energy e.g. Majorana modes in the Kitaev chain \cite{Kitaev1}; however, in non-Hermitian models this constrains energies according to $\text{Re}[E_{\pm}]=0$ which implies that the imaginary energy can be non-zero. Crucially, these modes cannot couple to bulk modes  as long as there is a gap in real energy and hence edge energies are pinned to be purely imaginary.

Can edge modes with $\text{Re}[E_{\pm}]=0$ be adiabatically removed without closing the real energy bulk band gap? The answer is no, as long as symmetries are respected and $H_{\text{BdG}}$ remains diagonalizable during the deformation process. In principle, the two modes with energies $\pm iu$ can become deformed to degeneracy at zero energy, allowing the associated eigenvectors to couple  via $t_{1,2}=t_{+} \pm i t_{-}$ such that the resulting modes $t_{1,2}$ are \textit{not} particle-hole symmetric \footnote{This can be achieved e.g. by adding a staggered real potential $ s \sum_{i=1}^{n}  \left(  b_{A,i}^{\dagger} b_{A,i}  - b_{B,i}^{\dagger} b_{B,i} \right)$ to the Hamiltonian \eqref{eq:bdgssh} and tuning from $s=0$ to $s = 2|u|$.}. However this process necessarily involves passing through an exceptional point since edge modes undergo a broken-to-unbroken transition at the degeneracy point (see Appendix) where $H_{\text{BdG}}$ becomes non-diagonalizable. This suggests that the diagonalizability of $H_{\text{BdG}}$ is important in the definition of a topological phase transition--a restriction which does not arise for Hermitian models and makes the physical protection weaker.

\section{Discussion and conclusions}

In this work, we have emphasized the need to go beyond the ten-fold way symmetry classes in order to generalize our understanding of symmetry-protected topological phase transitions. Our main motivation came from the dissipative bipartite quantum-walk model,  where chiral symmetry is broken while edge modes are protected by a uniquely non-Hermitian relationship. This led us to consider four symmetry relations originally proposed by BL \cite{Bernard1}. While these classes are naturally relevant  for non-Hermitian single-particle Hamiltonians \cite{Esaki1}, we have argued that they are  also useful for \textit{Hermitian} bosonic BdG models. These BdGs are diagonalized via a non-unitary transformation such that their spectrum is equivalent to the eigenvalues of a non-Hermitian matrix which belongs to one of the BL classes. We have studied a 1D bosonic model and commented on the robustness of edge instabilities.

An open task remains to fully classify all  bosonic BdG models \cite{Peano1}. Do bosonic BdGs inherit the same classification as their fermionic counterparts? We have noted that while both possess a particle-hole symmetry, the bosonic models  are  pseudo-Hermitian which will effect their classification. This should have implications for magnonic models \cite{Murakami1,Joshi1, Pollman1, Lawler1}, photonic systems \cite{Clerk1, Clerk2, Clerk3}, in addition to ultracold atoms \cite{Barnett1, Barnett2}, both with and without dynamical instabilities.

A more ambitious next step involves constructing a ``periodic table" from the  43 distinct non-Hermitian BL classes.  We  note that the BL classes can distinguish topological from trivial models in any dimension since they include the original ten Hermitian classes.

\section*{Acknowledgments}

I would like to thank Ryan Barnett and Derek~K.~K. Lee for inspiring  discussions. Funding from the Imperial College President's Scholarship is gratefully acknowledged.

\appendix

\section{\label{sec:apA} Consequences of pseudo-Hermiticity}

In this Appendix we discuss consequences of pseudo-Hermiticity since the spectrum of a bosonic BdG Hamiltonian is generically found by diagonalizing a pseudo-Hermitian matrix. This relation additionally plays an important role in formulations of non-Hermitian quantum mechanics \cite{Mostafazadeh1}.  Hermiticity ensures that all eigenvalues are real and right/left-eigenvectors are equivalent. A pseudo-Hermitian matrix is defined by the relation: $H = qH^{\dagger}q^{-1}$. Each eigenvalue of a non-Hermitian Hamiltonian has an associated  right and left eigenvector, written as:
\begin{align}
H \psi_{n}  = E_n  \psi_{n}  \label{eq:rightevec} \\
H^{\dagger} \lambda_{n}  = E_n^*  \lambda_{n} . \label{eq:leftevec}
\end{align}
Assuming pseudo-Hermiticity, if $\lambda_{n}  = q^{-1} \psi_{n} $, then we are guaranteed  $E_n \in \mathbb{R}$. Alternatively, if $\lambda_{n'}  = q^{-1} \psi_{n} $ then $E_n = E_{n'}^*$. So we find that eigenvalues are either purely real or come in conjugate pairs.

Suppose a pseudo-Hermitian Hamiltonian $H(\alpha)$ is parameterized by some value $\alpha$ such that all its eigenvalues are non-degenerate and real when $\alpha=0$. In order for complex modes to appear as we tune $\alpha$, two real eigenvalues must ``coalesce" (i.e. become degenerate) at $\alpha_c$ such that a complex conjugate pair forms adiabatically. The degeneracy point $\alpha_c$ is known as an exceptional point where the Hamiltonian is non-diagonalizable. This is also called a broken-to-unbroken transition point. It is easy to show that bulk modes undergo a broken-to-unbroken transition as a function of quasi-momentum $k$ for the parameter values $|w-v|<|u|$ for the bosonic-SSH. Additionally, pairs of modes on a given edge with energy $E_{\pm}=\pm i |u|$ can be deformed to have real energies via a broken-to-unbroken transition. 

\bibliography{classBib} 

\begin{thebibliography}{46}%
\makeatletter
\providecommand \@ifxundefined [1]{%
 \@ifx{#1\undefined}
}%
\providecommand \@ifnum [1]{%
 \ifnum #1\expandafter \@firstoftwo
 \else \expandafter \@secondoftwo
 \fi
}%
\providecommand \@ifx [1]{%
 \ifx #1\expandafter \@firstoftwo
 \else \expandafter \@secondoftwo
 \fi
}%
\providecommand \natexlab [1]{#1}%
\providecommand \enquote  [1]{``#1''}%
\providecommand \bibnamefont  [1]{#1}%
\providecommand \bibfnamefont [1]{#1}%
\providecommand \citenamefont [1]{#1}%
\providecommand \href@noop [0]{\@secondoftwo}%
\providecommand \href [0]{\begingroup \@sanitize@url \@href}%
\providecommand \@href[1]{\@@startlink{#1}\@@href}%
\providecommand \@@href[1]{\endgroup#1\@@endlink}%
\providecommand \@sanitize@url [0]{\catcode `\\12\catcode `\$12\catcode
  `\&12\catcode `\#12\catcode `\^12\catcode `\_12\catcode `\%12\relax}%
\providecommand \@@startlink[1]{}%
\providecommand \@@endlink[0]{}%
\providecommand \url  [0]{\begingroup\@sanitize@url \@url }%
\providecommand \@url [1]{\endgroup\@href {#1}{\urlprefix }}%
\providecommand \urlprefix  [0]{URL }%
\providecommand \Eprint [0]{\href }%
\providecommand \doibase [0]{http://dx.doi.org/}%
\providecommand \selectlanguage [0]{\@gobble}%
\providecommand \bibinfo  [0]{\@secondoftwo}%
\providecommand \bibfield  [0]{\@secondoftwo}%
\providecommand \translation [1]{[#1]}%
\providecommand \BibitemOpen [0]{}%
\providecommand \bibitemStop [0]{}%
\providecommand \bibitemNoStop [0]{.\EOS\space}%
\providecommand \EOS [0]{\spacefactor3000\relax}%
\providecommand \BibitemShut  [1]{\csname bibitem#1\endcsname}%
\let\auto@bib@innerbib\@empty
\bibitem [{\citenamefont {Qi}\ and\ \citenamefont {Zhang}(2011)}]{Qi1}%
  \BibitemOpen
  \bibfield  {author} {\bibinfo {author} {\bibfnamefont {X.-L.}\ \bibnamefont
  {Qi}}\ and\ \bibinfo {author} {\bibfnamefont {S.-C.}\ \bibnamefont {Zhang}},\
  }\href {\doibase 10.1103/RevModPhys.83.1057} {\bibfield  {journal} {\bibinfo
  {journal} {Rev. Mod. Phys.}\ }\textbf {\bibinfo {volume} {83}},\ \bibinfo
  {pages} {1057} (\bibinfo {year} {2011})}\BibitemShut {NoStop}%
\bibitem [{\citenamefont {Hasan}\ and\ \citenamefont {Kane}(2010)}]{Hasan1}%
  \BibitemOpen
  \bibfield  {author} {\bibinfo {author} {\bibfnamefont {M.~Z.}\ \bibnamefont
  {Hasan}}\ and\ \bibinfo {author} {\bibfnamefont {C.~L.}\ \bibnamefont
  {Kane}},\ }\href {\doibase 10.1103/RevModPhys.82.3045} {\bibfield  {journal}
  {\bibinfo  {journal} {Rev. Mod. Phys.}\ }\textbf {\bibinfo {volume} {82}},\
  \bibinfo {pages} {3045} (\bibinfo {year} {2010})}\BibitemShut {NoStop}%
\bibitem [{\citenamefont {Weimann}\ \emph {et~al.}(2017)\citenamefont
  {Weimann}, \citenamefont {Kremer}, \citenamefont {Plotnik}, \citenamefont
  {Lumer}, \citenamefont {Nolte}, \citenamefont {Makris}, \citenamefont
  {Segev}, \citenamefont {Rechtsman},\ and\ \citenamefont
  {Szameit}}]{Weimann1}%
  \BibitemOpen
  \bibfield  {author} {\bibinfo {author} {\bibfnamefont {A.}~\bibnamefont
  {Weimann}}, \bibinfo {author} {\bibfnamefont {M.}~\bibnamefont {Kremer}},
  \bibinfo {author} {\bibfnamefont {Y.}~\bibnamefont {Plotnik}}, \bibinfo
  {author} {\bibfnamefont {Y.}~\bibnamefont {Lumer}}, \bibinfo {author}
  {\bibfnamefont {S.}~\bibnamefont {Nolte}}, \bibinfo {author} {\bibfnamefont
  {K.}~\bibnamefont {Makris}}, \bibinfo {author} {\bibfnamefont
  {M.}~\bibnamefont {Segev}}, \bibinfo {author} {\bibfnamefont
  {M.}~\bibnamefont {Rechtsman}}, \ and\ \bibinfo {author} {\bibfnamefont
  {A.}~\bibnamefont {Szameit}},\ }\href {\doibase doi:10.1038/nmat4811}
  {\bibfield  {journal} {\bibinfo  {journal} {Nature materials}\ }\textbf
  {\bibinfo {volume} {16}},\ \bibinfo {pages} {433} (\bibinfo {year}
  {2017})}\BibitemShut {NoStop}%
\bibitem [{\citenamefont {Poli}\ \emph {et~al.}(2015)\citenamefont {Poli},
  \citenamefont {Bellec}, \citenamefont {Kuhl}, \citenamefont {Mortessagne},\
  and\ \citenamefont {Schomerus}}]{Poli1}%
  \BibitemOpen
  \bibfield  {author} {\bibinfo {author} {\bibfnamefont {C.}~\bibnamefont
  {Poli}}, \bibinfo {author} {\bibfnamefont {M.}~\bibnamefont {Bellec}},
  \bibinfo {author} {\bibfnamefont {U.}~\bibnamefont {Kuhl}}, \bibinfo {author}
  {\bibfnamefont {F.}~\bibnamefont {Mortessagne}}, \ and\ \bibinfo {author}
  {\bibfnamefont {H.}~\bibnamefont {Schomerus}},\ }\href {\doibase
  10.1038/ncomms7710 (2015)} {\bibfield  {journal} {\bibinfo  {journal} {Nature
  communications}\ }\textbf {\bibinfo {volume} {6}},\ \bibinfo {pages} {6710}
  (\bibinfo {year} {2015})}\BibitemShut {NoStop}%
\bibitem [{\citenamefont {Xiao}\ \emph {et~al.}(2017)\citenamefont {Xiao},
  \citenamefont {Zhan}, \citenamefont {Bian}, \citenamefont {Wang},
  \citenamefont {Zhang}, \citenamefont {Wang}, \citenamefont {Li},
  \citenamefont {Mochizuki}, \citenamefont {Kim}, \citenamefont {Kawakami},
  \citenamefont {Yi}, \citenamefont {Obuse}, \citenamefont {Sanders},\ and\
  \citenamefont {Xue}}]{Peng1}%
  \BibitemOpen
  \bibfield  {author} {\bibinfo {author} {\bibfnamefont {L.}~\bibnamefont
  {Xiao}}, \bibinfo {author} {\bibfnamefont {X.}~\bibnamefont {Zhan}}, \bibinfo
  {author} {\bibfnamefont {Z.~H.}\ \bibnamefont {Bian}}, \bibinfo {author}
  {\bibfnamefont {K.~K.}\ \bibnamefont {Wang}}, \bibinfo {author}
  {\bibfnamefont {X.}~\bibnamefont {Zhang}}, \bibinfo {author} {\bibfnamefont
  {X.~P.}\ \bibnamefont {Wang}}, \bibinfo {author} {\bibfnamefont
  {J.}~\bibnamefont {Li}}, \bibinfo {author} {\bibfnamefont {K.}~\bibnamefont
  {Mochizuki}}, \bibinfo {author} {\bibfnamefont {D.}~\bibnamefont {Kim}},
  \bibinfo {author} {\bibfnamefont {N.}~\bibnamefont {Kawakami}}, \bibinfo
  {author} {\bibfnamefont {W.}~\bibnamefont {Yi}}, \bibinfo {author}
  {\bibfnamefont {H.}~\bibnamefont {Obuse}}, \bibinfo {author} {\bibfnamefont
  {B.~C.}\ \bibnamefont {Sanders}}, \ and\ \bibinfo {author} {\bibfnamefont
  {P.}~\bibnamefont {Xue}},\ }\href {http://dx.doi.org/10.1038/nphys4204}
  {\bibfield  {journal} {\bibinfo  {journal} {Nature Physics}\ }\textbf
  {\bibinfo {volume} {13}},\ \bibinfo {pages} {1117 EP } (\bibinfo {year}
  {2017})}\BibitemShut {NoStop}%
\bibitem [{\citenamefont {Pocock}\ \emph {et~al.}()\citenamefont {Pocock},
  \citenamefont {Xiao}, \citenamefont {Huidobro},\ and\ \citenamefont
  {Giannini}}]{Pocock1}%
  \BibitemOpen
  \bibfield  {author} {\bibinfo {author} {\bibfnamefont {S.~R.}\ \bibnamefont
  {Pocock}}, \bibinfo {author} {\bibfnamefont {X.}~\bibnamefont {Xiao}},
  \bibinfo {author} {\bibfnamefont {P.~A.}\ \bibnamefont {Huidobro}}, \ and\
  \bibinfo {author} {\bibfnamefont {V.}~\bibnamefont {Giannini}},\ }\href
  {\doibase 10.1021/acsphotonics.8b00117} {\bibfield  {journal} {\bibinfo
  {journal} {ACS Photonics}\ }10.1021/acsphotonics.8b00117}\BibitemShut
  {NoStop}%
\bibitem [{\citenamefont {Downing}\ and\ \citenamefont
  {Weick}(2018)}]{Downing1}%
  \BibitemOpen
  \bibfield  {author} {\bibinfo {author} {\bibfnamefont {C.~A.}\ \bibnamefont
  {Downing}}\ and\ \bibinfo {author} {\bibfnamefont {G.}~\bibnamefont
  {Weick}},\ }\href@noop {} {\bibfield  {journal} {\bibinfo  {journal}
  {arXiv:1803.08872}\ } (\bibinfo {year} {2018})}\BibitemShut {NoStop}%
\bibitem [{\citenamefont {St-Jean}\ \emph {et~al.}(2017)\citenamefont
  {St-Jean}, \citenamefont {Goblot}, \citenamefont {Galopin}, \citenamefont
  {Lema{\^{\i}}tre}, \citenamefont {Ozawa}, \citenamefont {Gratiet},
  \citenamefont {Sagnes}, \citenamefont {Bloch},\ and\ \citenamefont
  {Amo}}]{Amo1}%
  \BibitemOpen
  \bibfield  {author} {\bibinfo {author} {\bibfnamefont {P.}~\bibnamefont
  {St-Jean}}, \bibinfo {author} {\bibfnamefont {V.}~\bibnamefont {Goblot}},
  \bibinfo {author} {\bibfnamefont {E.}~\bibnamefont {Galopin}}, \bibinfo
  {author} {\bibfnamefont {A.}~\bibnamefont {Lema{\^{\i}}tre}}, \bibinfo
  {author} {\bibfnamefont {T.}~\bibnamefont {Ozawa}}, \bibinfo {author}
  {\bibfnamefont {L.~L.}\ \bibnamefont {Gratiet}}, \bibinfo {author}
  {\bibfnamefont {I.}~\bibnamefont {Sagnes}}, \bibinfo {author} {\bibfnamefont
  {J.}~\bibnamefont {Bloch}}, \ and\ \bibinfo {author} {\bibfnamefont
  {A.}~\bibnamefont {Amo}},\ }\href {\doibase 10.1038/s41566-017-0006-2}
  {\bibfield  {journal} {\bibinfo  {journal} {Nature Photonics}\ }\textbf
  {\bibinfo {volume} {11}},\ \bibinfo {pages} {651} (\bibinfo {year}
  {2017})}\BibitemShut {NoStop}%
\bibitem [{\citenamefont {Malzard}\ \emph {et~al.}(2015)\citenamefont
  {Malzard}, \citenamefont {Poli},\ and\ \citenamefont {Schomerus}}]{Malzard1}%
  \BibitemOpen
  \bibfield  {author} {\bibinfo {author} {\bibfnamefont {S.}~\bibnamefont
  {Malzard}}, \bibinfo {author} {\bibfnamefont {C.}~\bibnamefont {Poli}}, \
  and\ \bibinfo {author} {\bibfnamefont {H.}~\bibnamefont {Schomerus}},\ }\href
  {\doibase 10.1103/PhysRevLett.115.200402} {\bibfield  {journal} {\bibinfo
  {journal} {Phys. Rev. Lett.}\ }\textbf {\bibinfo {volume} {115}},\ \bibinfo
  {pages} {200402} (\bibinfo {year} {2015})}\BibitemShut {NoStop}%
\bibitem [{\citenamefont {Kozii}\ and\ \citenamefont {Fu}(2017)}]{Kozii1}%
  \BibitemOpen
  \bibfield  {author} {\bibinfo {author} {\bibfnamefont {V.}~\bibnamefont
  {Kozii}}\ and\ \bibinfo {author} {\bibfnamefont {L.}~\bibnamefont {Fu}},\
  }\href@noop {} {\bibfield  {journal} {\bibinfo  {journal} {arXiv:1708.05841}\
  } (\bibinfo {year} {2017})}\BibitemShut {NoStop}%
\bibitem [{\citenamefont {Shen}\ \emph {et~al.}(2018)\citenamefont {Shen},
  \citenamefont {Zhen},\ and\ \citenamefont {Fu}}]{Shen1}%
  \BibitemOpen
  \bibfield  {author} {\bibinfo {author} {\bibfnamefont {H.}~\bibnamefont
  {Shen}}, \bibinfo {author} {\bibfnamefont {B.}~\bibnamefont {Zhen}}, \ and\
  \bibinfo {author} {\bibfnamefont {L.}~\bibnamefont {Fu}},\ }\href {\doibase
  10.1103/PhysRevLett.120.146402} {\bibfield  {journal} {\bibinfo  {journal}
  {Phys. Rev. Lett.}\ }\textbf {\bibinfo {volume} {120}},\ \bibinfo {pages}
  {146402} (\bibinfo {year} {2018})}\BibitemShut {NoStop}%
\bibitem [{\citenamefont {Gong}\ \emph {et~al.}(2018)\citenamefont {Gong},
  \citenamefont {Ashida}, \citenamefont {Kawabata}, \citenamefont {Takasan},
  \citenamefont {Higashikawa},\ and\ \citenamefont {Ueda}}]{Gong1}%
  \BibitemOpen
  \bibfield  {author} {\bibinfo {author} {\bibfnamefont {Z.}~\bibnamefont
  {Gong}}, \bibinfo {author} {\bibfnamefont {Y.}~\bibnamefont {Ashida}},
  \bibinfo {author} {\bibfnamefont {K.}~\bibnamefont {Kawabata}}, \bibinfo
  {author} {\bibfnamefont {K.}~\bibnamefont {Takasan}}, \bibinfo {author}
  {\bibfnamefont {S.}~\bibnamefont {Higashikawa}}, \ and\ \bibinfo {author}
  {\bibfnamefont {M.}~\bibnamefont {Ueda}},\ }\href@noop {} {\bibfield
  {journal} {\bibinfo  {journal} {arXiv:1802.07964}\ } (\bibinfo {year}
  {2018})}\BibitemShut {NoStop}%
\bibitem [{\citenamefont {Lee}(2016)}]{Lee1}%
  \BibitemOpen
  \bibfield  {author} {\bibinfo {author} {\bibfnamefont {T.~E.}\ \bibnamefont
  {Lee}},\ }\href {\doibase 10.1103/PhysRevLett.116.133903} {\bibfield
  {journal} {\bibinfo  {journal} {Phys. Rev. Lett.}\ }\textbf {\bibinfo
  {volume} {116}},\ \bibinfo {pages} {133903} (\bibinfo {year}
  {2016})}\BibitemShut {NoStop}%
\bibitem [{\citenamefont {Zhou}\ \emph {et~al.}(2018)\citenamefont {Zhou},
  \citenamefont {Peng}, \citenamefont {Yoon}, \citenamefont {Hsu},
  \citenamefont {Nelson}, \citenamefont {Fu}, \citenamefont {Joannopoulos},
  \citenamefont {Solja{\v c}i{\'c}},\ and\ \citenamefont {Zhen}}]{Zhou1}%
  \BibitemOpen
  \bibfield  {author} {\bibinfo {author} {\bibfnamefont {H.}~\bibnamefont
  {Zhou}}, \bibinfo {author} {\bibfnamefont {C.}~\bibnamefont {Peng}}, \bibinfo
  {author} {\bibfnamefont {Y.}~\bibnamefont {Yoon}}, \bibinfo {author}
  {\bibfnamefont {C.~W.}\ \bibnamefont {Hsu}}, \bibinfo {author} {\bibfnamefont
  {K.~A.}\ \bibnamefont {Nelson}}, \bibinfo {author} {\bibfnamefont
  {L.}~\bibnamefont {Fu}}, \bibinfo {author} {\bibfnamefont {J.~D.}\
  \bibnamefont {Joannopoulos}}, \bibinfo {author} {\bibfnamefont
  {M.}~\bibnamefont {Solja{\v c}i{\'c}}}, \ and\ \bibinfo {author}
  {\bibfnamefont {B.}~\bibnamefont {Zhen}},\ }\href {\doibase
  10.1126/science.aap9859} {\bibfield  {journal} {\bibinfo  {journal}
  {Science}\ } (\bibinfo {year} {2018}),\ 10.1126/science.aap9859}\BibitemShut
  {NoStop}%
\bibitem [{\citenamefont {Kunst}\ \emph {et~al.}(2018)\citenamefont {Kunst},
  \citenamefont {Edvardsson}, \citenamefont {Budich},\ and\ \citenamefont
  {Bergholtz}}]{Kunst1}%
  \BibitemOpen
  \bibfield  {author} {\bibinfo {author} {\bibfnamefont {F.~K.}\ \bibnamefont
  {Kunst}}, \bibinfo {author} {\bibfnamefont {E.}~\bibnamefont {Edvardsson}},
  \bibinfo {author} {\bibfnamefont {J.~C.}\ \bibnamefont {Budich}}, \ and\
  \bibinfo {author} {\bibfnamefont {E.~J.}\ \bibnamefont {Bergholtz}},\
  }\href@noop {} {\bibfield  {journal} {\bibinfo  {journal} {arXiv:1805.06492}\
  } (\bibinfo {year} {2018})}\BibitemShut {NoStop}%
\bibitem [{\citenamefont {Yao}\ and\ \citenamefont {Wang}(2018)}]{Yao1}%
  \BibitemOpen
  \bibfield  {author} {\bibinfo {author} {\bibfnamefont {S.}~\bibnamefont
  {Yao}}\ and\ \bibinfo {author} {\bibfnamefont {Z.}~\bibnamefont {Wang}},\
  }\href@noop {} {\bibfield  {journal} {\bibinfo  {journal} {arXiv:1803.01876}\
  } (\bibinfo {year} {2018})}\BibitemShut {NoStop}%
\bibitem [{\citenamefont {Yin}\ \emph {et~al.}(2018)\citenamefont {Yin},
  \citenamefont {Jiang}, \citenamefont {Li}, \citenamefont {L\"u},\ and\
  \citenamefont {Chen}}]{Yin1}%
  \BibitemOpen
  \bibfield  {author} {\bibinfo {author} {\bibfnamefont {C.}~\bibnamefont
  {Yin}}, \bibinfo {author} {\bibfnamefont {H.}~\bibnamefont {Jiang}}, \bibinfo
  {author} {\bibfnamefont {L.}~\bibnamefont {Li}}, \bibinfo {author}
  {\bibfnamefont {R.}~\bibnamefont {L\"u}}, \ and\ \bibinfo {author}
  {\bibfnamefont {S.}~\bibnamefont {Chen}},\ }\href {\doibase
  10.1103/PhysRevA.97.052115} {\bibfield  {journal} {\bibinfo  {journal} {Phys.
  Rev. A}\ }\textbf {\bibinfo {volume} {97}},\ \bibinfo {pages} {052115}
  (\bibinfo {year} {2018})}\BibitemShut {NoStop}%
\bibitem [{\citenamefont {Klett}\ \emph {et~al.}(2017)\citenamefont {Klett},
  \citenamefont {Cartarius}, \citenamefont {Dast}, \citenamefont {Main},\ and\
  \citenamefont {Wunner}}]{Wunner1}%
  \BibitemOpen
  \bibfield  {author} {\bibinfo {author} {\bibfnamefont {M.}~\bibnamefont
  {Klett}}, \bibinfo {author} {\bibfnamefont {H.}~\bibnamefont {Cartarius}},
  \bibinfo {author} {\bibfnamefont {D.}~\bibnamefont {Dast}}, \bibinfo {author}
  {\bibfnamefont {J.}~\bibnamefont {Main}}, \ and\ \bibinfo {author}
  {\bibfnamefont {G.}~\bibnamefont {Wunner}},\ }\href {\doibase
  10.1103/PhysRevA.95.053626} {\bibfield  {journal} {\bibinfo  {journal} {Phys.
  Rev. A}\ }\textbf {\bibinfo {volume} {95}},\ \bibinfo {pages} {053626}
  (\bibinfo {year} {2017})}\BibitemShut {NoStop}%
\bibitem [{\citenamefont {Lieu}(2018)}]{Lieu}%
  \BibitemOpen
  \bibfield  {author} {\bibinfo {author} {\bibfnamefont {S.}~\bibnamefont
  {Lieu}},\ }\href {\doibase 10.1103/PhysRevB.97.045106} {\bibfield  {journal}
  {\bibinfo  {journal} {Phys. Rev. B}\ }\textbf {\bibinfo {volume} {97}},\
  \bibinfo {pages} {045106} (\bibinfo {year} {2018})}\BibitemShut {NoStop}%
\bibitem [{\citenamefont {Altland}\ and\ \citenamefont
  {Zirnbauer}(1997)}]{Altland1}%
  \BibitemOpen
  \bibfield  {author} {\bibinfo {author} {\bibfnamefont {A.}~\bibnamefont
  {Altland}}\ and\ \bibinfo {author} {\bibfnamefont {M.~R.}\ \bibnamefont
  {Zirnbauer}},\ }\href {\doibase 10.1103/PhysRevB.55.1142} {\bibfield
  {journal} {\bibinfo  {journal} {Phys. Rev. B}\ }\textbf {\bibinfo {volume}
  {55}},\ \bibinfo {pages} {1142} (\bibinfo {year} {1997})}\BibitemShut
  {NoStop}%
\bibitem [{\citenamefont {Ryu}\ \emph {et~al.}(2010)\citenamefont {Ryu},
  \citenamefont {Schnyder}, \citenamefont {Furusaki},\ and\ \citenamefont
  {Ludwig}}]{Ryu1}%
  \BibitemOpen
  \bibfield  {author} {\bibinfo {author} {\bibfnamefont {S.}~\bibnamefont
  {Ryu}}, \bibinfo {author} {\bibfnamefont {A.~P.}\ \bibnamefont {Schnyder}},
  \bibinfo {author} {\bibfnamefont {A.}~\bibnamefont {Furusaki}}, \ and\
  \bibinfo {author} {\bibfnamefont {A.~W.~W.}\ \bibnamefont {Ludwig}},\ }\href
  {http://stacks.iop.org/1367-2630/12/i=6/a=065010} {\bibfield  {journal}
  {\bibinfo  {journal} {New Journal of Physics}\ }\textbf {\bibinfo {volume}
  {12}},\ \bibinfo {pages} {065010} (\bibinfo {year} {2010})}\BibitemShut
  {NoStop}%
\bibitem [{\citenamefont {Bernard}\ and\ \citenamefont
  {LeClair}(2002)}]{Bernard1}%
  \BibitemOpen
  \bibfield  {author} {\bibinfo {author} {\bibfnamefont {D.}~\bibnamefont
  {Bernard}}\ and\ \bibinfo {author} {\bibfnamefont {A.}~\bibnamefont
  {LeClair}},\ }\enquote {\bibinfo {title} {A classification of non-hermitian
  random matrices},}\ in\ \href {\doibase 10.1007/978-94-010-0514-2_19} {\emph
  {\bibinfo {booktitle} {Statistical Field Theories}}},\ \bibinfo {editor}
  {edited by\ \bibinfo {editor} {\bibfnamefont {A.}~\bibnamefont {Cappelli}}\
  and\ \bibinfo {editor} {\bibfnamefont {G.}~\bibnamefont {Mussardo}}}\
  (\bibinfo  {publisher} {Springer Netherlands},\ \bibinfo {address}
  {Dordrecht},\ \bibinfo {year} {2002})\ pp.\ \bibinfo {pages}
  {207--214}\BibitemShut {NoStop}%
\bibitem [{\citenamefont {Esaki}\ \emph {et~al.}(2011)\citenamefont {Esaki},
  \citenamefont {Sato}, \citenamefont {Hasebe},\ and\ \citenamefont
  {Kohmoto}}]{Esaki1}%
  \BibitemOpen
  \bibfield  {author} {\bibinfo {author} {\bibfnamefont {K.}~\bibnamefont
  {Esaki}}, \bibinfo {author} {\bibfnamefont {M.}~\bibnamefont {Sato}},
  \bibinfo {author} {\bibfnamefont {K.}~\bibnamefont {Hasebe}}, \ and\ \bibinfo
  {author} {\bibfnamefont {M.}~\bibnamefont {Kohmoto}},\ }\href {\doibase
  10.1103/PhysRevB.84.205128} {\bibfield  {journal} {\bibinfo  {journal} {Phys.
  Rev. B}\ }\textbf {\bibinfo {volume} {84}},\ \bibinfo {pages} {205128}
  (\bibinfo {year} {2011})}\BibitemShut {NoStop}%
\bibitem [{\citenamefont {Gurarie}\ and\ \citenamefont
  {Chalker}(2003)}]{Chalker1}%
  \BibitemOpen
  \bibfield  {author} {\bibinfo {author} {\bibfnamefont {V.}~\bibnamefont
  {Gurarie}}\ and\ \bibinfo {author} {\bibfnamefont {J.~T.}\ \bibnamefont
  {Chalker}},\ }\href {\doibase 10.1103/PhysRevB.68.134207} {\bibfield
  {journal} {\bibinfo  {journal} {Phys. Rev. B}\ }\textbf {\bibinfo {volume}
  {68}},\ \bibinfo {pages} {134207} (\bibinfo {year} {2003})}\BibitemShut
  {NoStop}%
\bibitem [{\citenamefont {Miesner}\ \emph {et~al.}(1999)\citenamefont
  {Miesner}, \citenamefont {Stamper-Kurn}, \citenamefont {Stenger},
  \citenamefont {Inouye}, \citenamefont {Chikkatur},\ and\ \citenamefont
  {Ketterle}}]{Ketterle1}%
  \BibitemOpen
  \bibfield  {author} {\bibinfo {author} {\bibfnamefont {H.-J.}\ \bibnamefont
  {Miesner}}, \bibinfo {author} {\bibfnamefont {D.~M.}\ \bibnamefont
  {Stamper-Kurn}}, \bibinfo {author} {\bibfnamefont {J.}~\bibnamefont
  {Stenger}}, \bibinfo {author} {\bibfnamefont {S.}~\bibnamefont {Inouye}},
  \bibinfo {author} {\bibfnamefont {A.~P.}\ \bibnamefont {Chikkatur}}, \ and\
  \bibinfo {author} {\bibfnamefont {W.}~\bibnamefont {Ketterle}},\ }\href
  {\doibase 10.1103/PhysRevLett.82.2228} {\bibfield  {journal} {\bibinfo
  {journal} {Phys. Rev. Lett.}\ }\textbf {\bibinfo {volume} {82}},\ \bibinfo
  {pages} {2228} (\bibinfo {year} {1999})}\BibitemShut {NoStop}%
\bibitem [{\citenamefont {Bernier}\ \emph {et~al.}(2014)\citenamefont
  {Bernier}, \citenamefont {Dalla~Torre},\ and\ \citenamefont
  {Demler}}]{Demler1}%
  \BibitemOpen
  \bibfield  {author} {\bibinfo {author} {\bibfnamefont {N.~R.}\ \bibnamefont
  {Bernier}}, \bibinfo {author} {\bibfnamefont {E.~G.}\ \bibnamefont
  {Dalla~Torre}}, \ and\ \bibinfo {author} {\bibfnamefont {E.}~\bibnamefont
  {Demler}},\ }\href {\doibase 10.1103/PhysRevLett.113.065303} {\bibfield
  {journal} {\bibinfo  {journal} {Phys. Rev. Lett.}\ }\textbf {\bibinfo
  {volume} {113}},\ \bibinfo {pages} {065303} (\bibinfo {year}
  {2014})}\BibitemShut {NoStop}%
\bibitem [{\citenamefont {Nakamura}\ \emph {et~al.}(2008)\citenamefont
  {Nakamura}, \citenamefont {Mine}, \citenamefont {Okumura},\ and\
  \citenamefont {Yamanaka}}]{Nakamura1}%
  \BibitemOpen
  \bibfield  {author} {\bibinfo {author} {\bibfnamefont {Y.}~\bibnamefont
  {Nakamura}}, \bibinfo {author} {\bibfnamefont {M.}~\bibnamefont {Mine}},
  \bibinfo {author} {\bibfnamefont {M.}~\bibnamefont {Okumura}}, \ and\
  \bibinfo {author} {\bibfnamefont {Y.}~\bibnamefont {Yamanaka}},\ }\href
  {\doibase 10.1103/PhysRevA.77.043601} {\bibfield  {journal} {\bibinfo
  {journal} {Phys. Rev. A}\ }\textbf {\bibinfo {volume} {77}},\ \bibinfo
  {pages} {043601} (\bibinfo {year} {2008})}\BibitemShut {NoStop}%
\bibitem [{\citenamefont {Barnett}(2013)}]{Barnett1}%
  \BibitemOpen
  \bibfield  {author} {\bibinfo {author} {\bibfnamefont {R.}~\bibnamefont
  {Barnett}},\ }\href {\doibase 10.1103/PhysRevA.88.063631} {\bibfield
  {journal} {\bibinfo  {journal} {Phys. Rev. A}\ }\textbf {\bibinfo {volume}
  {88}},\ \bibinfo {pages} {063631} (\bibinfo {year} {2013})}\BibitemShut
  {NoStop}%
\bibitem [{\citenamefont {Galilo}\ \emph {et~al.}(2015)\citenamefont {Galilo},
  \citenamefont {Lee},\ and\ \citenamefont {Barnett}}]{Barnett2}%
  \BibitemOpen
  \bibfield  {author} {\bibinfo {author} {\bibfnamefont {B.}~\bibnamefont
  {Galilo}}, \bibinfo {author} {\bibfnamefont {D.~K.~K.}\ \bibnamefont {Lee}},
  \ and\ \bibinfo {author} {\bibfnamefont {R.}~\bibnamefont {Barnett}},\ }\href
  {\doibase 10.1103/PhysRevLett.115.245302} {\bibfield  {journal} {\bibinfo
  {journal} {Phys. Rev. Lett.}\ }\textbf {\bibinfo {volume} {115}},\ \bibinfo
  {pages} {245302} (\bibinfo {year} {2015})}\BibitemShut {NoStop}%
\bibitem [{\citenamefont {Su}\ \emph {et~al.}(1979)\citenamefont {Su},
  \citenamefont {Schrieffer},\ and\ \citenamefont {Heeger}}]{Su1}%
  \BibitemOpen
  \bibfield  {author} {\bibinfo {author} {\bibfnamefont {W.~P.}\ \bibnamefont
  {Su}}, \bibinfo {author} {\bibfnamefont {J.~R.}\ \bibnamefont {Schrieffer}},
  \ and\ \bibinfo {author} {\bibfnamefont {A.~J.}\ \bibnamefont {Heeger}},\
  }\href {\doibase 10.1103/PhysRevLett.42.1698} {\bibfield  {journal} {\bibinfo
   {journal} {Phys. Rev. Lett.}\ }\textbf {\bibinfo {volume} {42}},\ \bibinfo
  {pages} {1698} (\bibinfo {year} {1979})}\BibitemShut {NoStop}%
\bibitem [{\citenamefont {Rudner}\ and\ \citenamefont
  {Levitov}(2009)}]{Rudner1}%
  \BibitemOpen
  \bibfield  {author} {\bibinfo {author} {\bibfnamefont {M.~S.}\ \bibnamefont
  {Rudner}}\ and\ \bibinfo {author} {\bibfnamefont {L.~S.}\ \bibnamefont
  {Levitov}},\ }\href {\doibase 10.1103/PhysRevLett.102.065703} {\bibfield
  {journal} {\bibinfo  {journal} {Phys. Rev. Lett.}\ }\textbf {\bibinfo
  {volume} {102}},\ \bibinfo {pages} {065703} (\bibinfo {year}
  {2009})}\BibitemShut {NoStop}%
\bibitem [{\citenamefont {Zeuner}\ \emph {et~al.}(2015)\citenamefont {Zeuner},
  \citenamefont {Rechtsman}, \citenamefont {Plotnik}, \citenamefont {Lumer},
  \citenamefont {Nolte}, \citenamefont {Rudner}, \citenamefont {Segev},\ and\
  \citenamefont {Szameit}}]{Rudner2}%
  \BibitemOpen
  \bibfield  {author} {\bibinfo {author} {\bibfnamefont {J.~M.}\ \bibnamefont
  {Zeuner}}, \bibinfo {author} {\bibfnamefont {M.~C.}\ \bibnamefont
  {Rechtsman}}, \bibinfo {author} {\bibfnamefont {Y.}~\bibnamefont {Plotnik}},
  \bibinfo {author} {\bibfnamefont {Y.}~\bibnamefont {Lumer}}, \bibinfo
  {author} {\bibfnamefont {S.}~\bibnamefont {Nolte}}, \bibinfo {author}
  {\bibfnamefont {M.~S.}\ \bibnamefont {Rudner}}, \bibinfo {author}
  {\bibfnamefont {M.}~\bibnamefont {Segev}}, \ and\ \bibinfo {author}
  {\bibfnamefont {A.}~\bibnamefont {Szameit}},\ }\href {\doibase
  10.1103/PhysRevLett.115.040402} {\bibfield  {journal} {\bibinfo  {journal}
  {Phys. Rev. Lett.}\ }\textbf {\bibinfo {volume} {115}},\ \bibinfo {pages}
  {040402} (\bibinfo {year} {2015})}\BibitemShut {NoStop}%
\bibitem [{Note1()}]{Note1}%
  \BibitemOpen
  \bibinfo {note} {The model studied was slightly different from the
  dissipative quantum walk, however the argument remains valid.}\BibitemShut
  {Stop}%
\bibitem [{\citenamefont {Magnea}(2008)}]{Magnea1}%
  \BibitemOpen
  \bibfield  {author} {\bibinfo {author} {\bibfnamefont {U.}~\bibnamefont
  {Magnea}},\ }\href {http://stacks.iop.org/1751-8121/41/i=4/a=045203}
  {\bibfield  {journal} {\bibinfo  {journal} {Journal of Physics A:
  Mathematical and Theoretical}\ }\textbf {\bibinfo {volume} {41}},\ \bibinfo
  {pages} {045203} (\bibinfo {year} {2008})}\BibitemShut {NoStop}%
\bibitem [{\citenamefont {Kitaev}(2001)}]{Kitaev1}%
  \BibitemOpen
  \bibfield  {author} {\bibinfo {author} {\bibfnamefont {A.~Y.}\ \bibnamefont
  {Kitaev}},\ }\href {http://stacks.iop.org/1063-7869/44/i=10S/a=S29}
  {\bibfield  {journal} {\bibinfo  {journal} {Physics-Uspekhi}\ }\textbf
  {\bibinfo {volume} {44}},\ \bibinfo {pages} {131} (\bibinfo {year}
  {2001})}\BibitemShut {NoStop}%
\bibitem [{\citenamefont {Takahashi}\ and\ \citenamefont
  {Nitta}(2015)}]{Takahashi1}%
  \BibitemOpen
  \bibfield  {author} {\bibinfo {author} {\bibfnamefont {D.~A.}\ \bibnamefont
  {Takahashi}}\ and\ \bibinfo {author} {\bibfnamefont {M.}~\bibnamefont
  {Nitta}},\ }\href {\doibase https://doi.org/10.1016/j.aop.2014.12.009}
  {\bibfield  {journal} {\bibinfo  {journal} {Annals of Physics}\ }\textbf
  {\bibinfo {volume} {354}},\ \bibinfo {pages} {101 } (\bibinfo {year}
  {2015})}\BibitemShut {NoStop}%
\bibitem [{Note2()}]{Note2}%
  \BibitemOpen
  \bibinfo {note} {This can be achieved e.g. by adding a staggered real
  potential $ s \DOTSB \sum@ \slimits@ _{i=1}^{n} \left ( b_{A,i}^{\dagger }
  b_{A,i} - b_{B,i}^{\dagger } b_{B,i} \right )$ to the Hamiltonian \protect
  \textup {\hbox {\mathsurround \z@ \protect \normalfont (\ignorespaces \ref
  {eq:bdgssh}\unskip \@@italiccorr )}} and tuning from $s=0$ to $s =
  2|u|$.}\BibitemShut {Stop}%
\bibitem [{\citenamefont {Peano}\ and\ \citenamefont
  {Schulz-Baldes}(2018)}]{Peano1}%
  \BibitemOpen
  \bibfield  {author} {\bibinfo {author} {\bibfnamefont {V.}~\bibnamefont
  {Peano}}\ and\ \bibinfo {author} {\bibfnamefont {H.}~\bibnamefont
  {Schulz-Baldes}},\ }\href {\doibase 10.1063/1.5002094} {\bibfield  {journal}
  {\bibinfo  {journal} {Journal of Mathematical Physics}\ }\textbf {\bibinfo
  {volume} {59}},\ \bibinfo {pages} {031901} (\bibinfo {year} {2018})},\
  \Eprint {http://arxiv.org/abs/https://doi.org/10.1063/1.5002094}
  {https://doi.org/10.1063/1.5002094} \BibitemShut {NoStop}%
\bibitem [{\citenamefont {Shindou}\ \emph {et~al.}(2013)\citenamefont
  {Shindou}, \citenamefont {Matsumoto}, \citenamefont {Murakami},\ and\
  \citenamefont {Ohe}}]{Murakami1}%
  \BibitemOpen
  \bibfield  {author} {\bibinfo {author} {\bibfnamefont {R.}~\bibnamefont
  {Shindou}}, \bibinfo {author} {\bibfnamefont {R.}~\bibnamefont {Matsumoto}},
  \bibinfo {author} {\bibfnamefont {S.}~\bibnamefont {Murakami}}, \ and\
  \bibinfo {author} {\bibfnamefont {J.-i.}\ \bibnamefont {Ohe}},\ }\href
  {\doibase 10.1103/PhysRevB.87.174427} {\bibfield  {journal} {\bibinfo
  {journal} {Phys. Rev. B}\ }\textbf {\bibinfo {volume} {87}},\ \bibinfo
  {pages} {174427} (\bibinfo {year} {2013})}\BibitemShut {NoStop}%
\bibitem [{\citenamefont {Joshi}(2018)}]{Joshi1}%
  \BibitemOpen
  \bibfield  {author} {\bibinfo {author} {\bibfnamefont {D.~G.}\ \bibnamefont
  {Joshi}},\ }\href {\doibase 10.1103/PhysRevB.98.060405} {\bibfield  {journal}
  {\bibinfo  {journal} {Phys. Rev. B}\ }\textbf {\bibinfo {volume} {98}},\
  \bibinfo {pages} {060405} (\bibinfo {year} {2018})}\BibitemShut {NoStop}%
\bibitem [{\citenamefont {McClarty}\ \emph {et~al.}(2018)\citenamefont
  {McClarty}, \citenamefont {Dong}, \citenamefont {Gohlke}, \citenamefont
  {Rau}, \citenamefont {Pollmann}, \citenamefont {Moessner},\ and\
  \citenamefont {Penc}}]{Pollman1}%
  \BibitemOpen
  \bibfield  {author} {\bibinfo {author} {\bibfnamefont {P.~A.}\ \bibnamefont
  {McClarty}}, \bibinfo {author} {\bibfnamefont {X.-Y.}\ \bibnamefont {Dong}},
  \bibinfo {author} {\bibfnamefont {M.}~\bibnamefont {Gohlke}}, \bibinfo
  {author} {\bibfnamefont {J.~G.}\ \bibnamefont {Rau}}, \bibinfo {author}
  {\bibfnamefont {F.}~\bibnamefont {Pollmann}}, \bibinfo {author}
  {\bibfnamefont {R.}~\bibnamefont {Moessner}}, \ and\ \bibinfo {author}
  {\bibfnamefont {K.}~\bibnamefont {Penc}},\ }\href {\doibase
  10.1103/PhysRevB.98.060404} {\bibfield  {journal} {\bibinfo  {journal} {Phys.
  Rev. B}\ }\textbf {\bibinfo {volume} {98}},\ \bibinfo {pages} {060404}
  (\bibinfo {year} {2018})}\BibitemShut {NoStop}%
\bibitem [{\citenamefont {Roychowdhury}\ and\ \citenamefont
  {Lawler}(2018)}]{Lawler1}%
  \BibitemOpen
  \bibfield  {author} {\bibinfo {author} {\bibfnamefont {K.}~\bibnamefont
  {Roychowdhury}}\ and\ \bibinfo {author} {\bibfnamefont {M.~J.}\ \bibnamefont
  {Lawler}},\ }\href@noop {} {\bibfield  {journal} {\bibinfo  {journal}
  {arXiv:1807.00837}\ } (\bibinfo {year} {2018})}\BibitemShut {NoStop}%
\bibitem [{\citenamefont {Peano}\ \emph
  {et~al.}(2016{\natexlab{a}})\citenamefont {Peano}, \citenamefont {Houde},
  \citenamefont {Marquardt},\ and\ \citenamefont {Clerk}}]{Clerk1}%
  \BibitemOpen
  \bibfield  {author} {\bibinfo {author} {\bibfnamefont {V.}~\bibnamefont
  {Peano}}, \bibinfo {author} {\bibfnamefont {M.}~\bibnamefont {Houde}},
  \bibinfo {author} {\bibfnamefont {F.}~\bibnamefont {Marquardt}}, \ and\
  \bibinfo {author} {\bibfnamefont {A.~A.}\ \bibnamefont {Clerk}},\ }\href
  {\doibase 10.1103/PhysRevX.6.041026} {\bibfield  {journal} {\bibinfo
  {journal} {Phys. Rev. X}\ }\textbf {\bibinfo {volume} {6}},\ \bibinfo {pages}
  {041026} (\bibinfo {year} {2016}{\natexlab{a}})}\BibitemShut {NoStop}%
\bibitem [{\citenamefont {Peano}\ \emph
  {et~al.}(2016{\natexlab{b}})\citenamefont {Peano}, \citenamefont {Houde},
  \citenamefont {Brendel}, \citenamefont {Marquardt},\ and\ \citenamefont
  {Clerk}}]{Clerk2}%
  \BibitemOpen
  \bibfield  {author} {\bibinfo {author} {\bibfnamefont {V.}~\bibnamefont
  {Peano}}, \bibinfo {author} {\bibfnamefont {M.}~\bibnamefont {Houde}},
  \bibinfo {author} {\bibfnamefont {C.}~\bibnamefont {Brendel}}, \bibinfo
  {author} {\bibfnamefont {F.}~\bibnamefont {Marquardt}}, \ and\ \bibinfo
  {author} {\bibfnamefont {A.~A.}\ \bibnamefont {Clerk}},\ }\href {\doibase
  10.1038/ncomms10779} {\bibfield  {journal} {\bibinfo  {journal} {Nature
  Communications}\ }\textbf {\bibinfo {volume} {7}},\ \bibinfo {pages} {10779}
  (\bibinfo {year} {2016}{\natexlab{b}})}\BibitemShut {NoStop}%
\bibitem [{\citenamefont {McDonald}\ \emph {et~al.}(2018)\citenamefont
  {McDonald}, \citenamefont {Pereg-Barnea},\ and\ \citenamefont
  {Clerk}}]{Clerk3}%
  \BibitemOpen
  \bibfield  {author} {\bibinfo {author} {\bibfnamefont {A.}~\bibnamefont
  {McDonald}}, \bibinfo {author} {\bibfnamefont {T.}~\bibnamefont
  {Pereg-Barnea}}, \ and\ \bibinfo {author} {\bibfnamefont {A.}~\bibnamefont
  {Clerk}},\ }\href@noop {} {\bibfield  {journal} {\bibinfo  {journal}
  {arXiv:1805.12557}\ } (\bibinfo {year} {2018})}\BibitemShut {NoStop}%
\bibitem [{\citenamefont {Mostafazadeh}(2002)}]{Mostafazadeh1}%
  \BibitemOpen
  \bibfield  {author} {\bibinfo {author} {\bibfnamefont {A.}~\bibnamefont
  {Mostafazadeh}},\ }\href {\doibase 10.1063/1.1418246} {\bibfield  {journal}
  {\bibinfo  {journal} {Journal of Mathematical Physics}\ }\textbf {\bibinfo
  {volume} {43}},\ \bibinfo {pages} {205} (\bibinfo {year} {2002})},\ \Eprint
  {http://arxiv.org/abs/https://doi.org/10.1063/1.1418246}
  {https://doi.org/10.1063/1.1418246} \BibitemShut {NoStop}%
\end{thebibliography}%
\bibliographystyle{apsrev4-1}

\end{document}